\documentclass[aps,prl,reprint,superscriptaddress,floatfix]{revtex4-2}

\usepackage{graphicx}
\usepackage{amsmath}
\usepackage{amssymb}
\usepackage{bm}
\usepackage{hyperref}
\usepackage{xcolor}

\definecolor{editgreen}{rgb}{0.0,0.5,0.0}
\definecolor{editred}{rgb}{0.8,0.0,0.0}

\begin{document}

\title{Optimal Persistence Reveals Hidden Topology in Complex Energy Landscapes}

\author{LI Zhenpeng}
\affiliation{School of Artificial Intelligence, Taizhou University, 318000 Taizhou Zhejiang Province, China}

\begin{abstract}
Infinite persistence marks the topological transition. 
For finite persistence, the canyon-finding rate $\Gamma(\tau_p)$ on the $p=2$ spherical spin glass forms an inverted-U profile, peaking at an optimal $\tau_p^*$.
At low temperature ($T=0.05$), $\tau_p^*$ drops from $10$ to $5$ as $N$ increases through $128$, marking the discrete-to-quasi-continuous GOE crossover. 
For $N=1024$, the peak is flat between $\tau_p=5$ and $6$ within statistical uncertainties, preventing a more precise determination.
For $N\ge128$, the canyon width saturates at $\xi_{\text{eff}}=1$, consistent with the measured $\tau_p^*=5$ when $\beta=0.4$. 
At higher temperatures ($T\ge0.15$), $\tau_p^*=10$ and $\beta(T)$ scales as $1/T$, with temperature dependence entering only through $v_{\text{th}}=\sqrt{2T}$. 
For $T=0.10$ and $N\ge128$, high-resolution scans give $\tau_p^*=8.0$; for $N\le64$ at the same temperature, coarse scans place $\tau_p^*$ in the range $8$--$10$.
Thus, optimal persistence reveals the hidden topology of the landscape — a principle expected to be generic in disordered landscapes with entropic bottlenecks.
\end{abstract}

\maketitle

\section{Introduction}

How does a persistent random walker explore a complex energy landscape? 
The accessible configuration space is not static but dynamically enacted by 
the walker's own trajectory and timescales \cite{sethna2021,wales2003energy}. 
Between the limits of zero and infinite persistence lies an optimal value — 
but is it set by system size, by temperature, or by the spectrum itself?

Kent-Dobias recently showed that in the limit of \textit{infinite} persistence, 
the walker's ergodicity-breaking point coincides exactly with the topological 
transition of the landscape's constant-energy slices \cite{kentdobias2026}. 
Persistence thus acts as a geometric detector. But that result concerns the 
boundary of ergodicity. What happens at \textit{finite} persistence? Is the 
approach monotonic, or does an intermediate, \textit{optimal} persistence 
maximize exploration efficiency?

In this Letter, we answer these questions by systematically studying a persistent 
Langevin walker on the $p=2$ spherical spin glass \cite{kosterlitz1976,crisanti1993} across two orders of magnitude 
in system size ($N=16$ to $1024$) and temperatures $T=0.05$ to $0.30$. 
We find that the route to infinite persistence is non-monotonic, revealing a 
sharp size-driven dynamical transition at low temperatures.

\subsection*{Main findings}
We uncover an inverted-U relationship between exploration efficiency $\Gamma$ 
and persistence time $\tau_p$ (Fig.~1). The optimal persistence $\tau_p^*$ 
exhibits a rich dependence on temperature and system size.

At high temperatures, $\tau_p^*$ is independent of $N$; at low temperature 
($T=0.05$), a sharp size-driven crossover occurs as $N$ increases through 
$128$ (Figs.~2 and~3). This crossover coincides with the discrete-to-
quasi-continuous transition in the GOE spectrum, where the entropic 
bottleneck width reaches its  maximum. 

At fixed $N=128$, $\tau_p^*$ decreases with decreasing temperature, 
revealing a thermal crossover that reflects geometric saturation of the 
bottleneck. These findings establish optimal persistence as a probe of 
hidden topology, with the optimal memory time set by the landscape's 
spectral statistics rather than by temperature alone.

The behavior in the unsaturated regime ($N<128$) remains an open problem 
for future work.

\subsection*{Physical interpretation}
The inverted-U follows from scanning efficiency ($\propto\tau_p$) versus 
trajectory sparseness ($\propto1/\tau_p$), yielding a Lorentzian form 
(End Matter Sec.~S2). The optimum corresponds to $\partial\Gamma/\partial E = 0$ 
at $e_c$ (End Matter Sec.~S8), where the Euler characteristic $\chi(E)$ 
exhibits a discontinuity \cite{kentdobias2026}. The $T=0.05$ crossover is 
observed between two regimes: discrete ($\tau_p^*=10$) and quasi-continuous 
($\tau_p^*=5$). Within the saturated regime ($N \ge 128$), the scaling is consistent with an RG fixed point. A full analytical 
description of the crossover between regimes remains an open question 
(End Matter Sec.~S6).

\subsection*{Broader implications}
This spectral-dynamical crossover reveals a distinct regime: in the 
quasi-continuous regime, $\tau_p^* = 5$ is independent of 
$\langle\Delta\lambda\rangle$, consistent with a $z=1$ dynamical exponent (see End Matter Sec.~S2)
and hinting at an infinite-randomness-like fixed point.

The principle of ``optimal persistence'' — matching the walker's memory 
time to the landscape's characteristic scale — extends beyond the 
$p=2$ spherical model. In active matter, colloids in porous media exhibit 
optimal activity set by geometric constraints \cite{kurzthaler2021geometric}. 
In optimization, stochastic gradient descent with momentum faces the same 
trade-off between jitter and overshoot. In glassy dynamics, the hierarchy 
of relaxation timescales may reflect spatially varying local bottlenecks. 
We conjecture that this spectral-dynamical crossover — optimal persistence 
governed by spectral statistics rather than geometry — is generic in 
disordered landscapes with entropic bottlenecks.

\section{Model and Methods}

\paragraph*{Energy Landscape}
We consider the spherical $p=2$ spin glass, whose energy landscape 
properties~\cite{kosterlitz1976,fyodorov2014} as well as Langevin dynamics 
at all scales~\cite{cugliandolo1995,benarous2001,fyodorov2015perret} are fully 
characterized by random matrix theory (RMT); see~\cite{fyodorov2005intro} 
for an accessible introduction to RMT.

The Hamiltonian is
\begin{equation}
H(\mathbf{x}) = -\frac{1}{2} \sum_{i,j=1}^N J_{ij} x_i x_j,
\label{eq:hamiltonian}
\end{equation}
with spherical constraint $\sum_i x_i^2 = N$. The couplings $J_{ij}$ are 
drawn from the Gaussian Orthogonal Ensemble (GOE) with zero mean and 
variance $\langle J_{ij}^2 \rangle = 1/N$. The largest eigenvalue 
$\lambda_{\max}$ determines the two symmetric global minima. At the 
critical energy $e_c = -\lambda_2/2$, where $\lambda_2$ is the 
second-largest eigenvalue, the equipotential surfaces undergo a 
topological transition. 
The spectral gap $\Delta\lambda = \lambda_{\max} - \lambda_2$ 
controls the width of the entropic bottleneck connecting the minima, 
and controls the final stages of gradient descent relaxation~\cite{fyodorov2015perret} 
(see End Matter Fig.~\ref{fig:S1})

\paragraph*{Persistent Dynamics}

The dynamic behavior of the $p=2$ spherical spin glass has been extensively studied 
\cite{crisanti1993,campellone1998}.
A persistent walker navigates this landscape via the underdamped 
Langevin equation projected onto the sphere:
\begin{equation}
m \ddot{\mathbf{x}} = -\gamma \dot{\mathbf{x}} - \nabla_{\!\perp} H(\mathbf{x}) 
+ \sqrt{2\gamma T} \, \boldsymbol{\eta}(t),
\label{eq:langevin}
\end{equation}
with constraints $|\mathbf{x}|^2 = N$ and $\mathbf{x} \cdot \dot{\mathbf{x}} = 0$. 
Here, $\nabla_{\!\perp}$ is the tangential gradient, $\boldsymbol{\eta}(t)$ 
is Gaussian white noise, and the persistence time $\tau_p = m/\gamma$ 
interpolates between overdamped ($\tau_p\to0$) and ballistic ($\tau_p\to\infty$) 
limits.

\paragraph*{Simulation Protocol}
We scan persistence times $\tau_p = 0.05$ to $50.0$, temperatures 
$T = 0.05$ to $0.30$, and system sizes $N = 16$ to $1024$. 
Walkers are initialized near the critical energy $e_c$.

The decay rate $\Gamma$—the canyon-finding rate—is extracted from exponential fits to the energy autocorrelation function $C_e(t)$. Full simulation parameters are listed in End Matter Table~\ref{tab:S2}.

\section{Results}

\subsection*{Resonance mechanism}

\begin{figure}[t]
\centering
\includegraphics[width=\columnwidth]{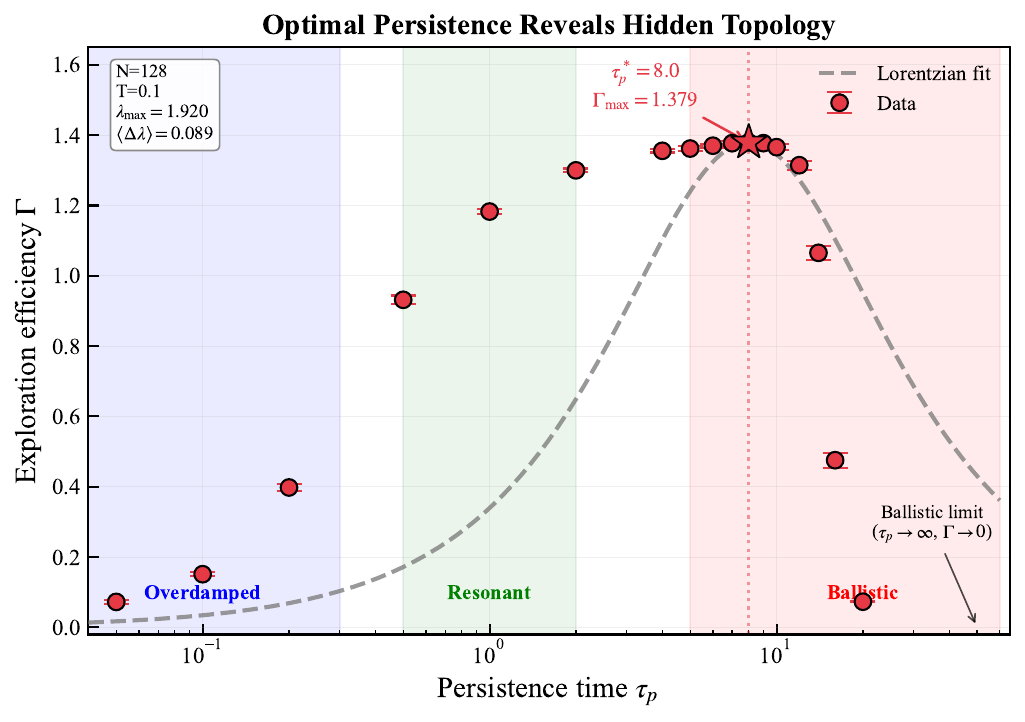}
\caption{Inverted-U profile of exploration efficiency for $N=128$ at $T=0.10$. 
Data points (red circles, error bars $\pm1\sigma$ from 500 realizations) with a Lorentzian fit (dashed gray curve): 
$\Gamma(\tau_p) = \Gamma_{\max}\cdot 2(\tau_p/\tau_p^*)/[1+(\tau_p/\tau_p^*)^2]$, 
$\tau_p^* = 8.0$, $\Gamma_{\max}=1.379$ (star). 
Three regimes: overdamped (blue, $\tau_p\ll\tau_p^*$), resonant (green, $\tau_p\sim\tau_p^*$), and ballistic (red, $\tau_p\gg\tau_p^*$). 
The ballistic limit $\Gamma\to0$ as $\tau_p\to\infty$ is indicated.}
\label{fig:main}
\end{figure}

Figure~\ref{fig:main} displays $\Gamma(\tau_p)$ for $N=128$ at $T=0.10$. 
A clear inverted-U emerges. The optimal persistence $\tau_p^* = 8.0$ marks 
the resonance where the walker's persistence length matches the entropic 
bottleneck width.

Three dynamical regimes are identified: overdamped ($\tau_p \lesssim 2.0$), 
where local jittering yields low efficiency ($\Gamma \lesssim 0.9$); resonant 
($\tau_p^* = 8.0$), where efficiency peaks at $\Gamma_{\max} = 1.379$; and 
ballistic ($\tau_p \gtrsim 20.0$), where geodesic trajectories bypass the 
channel and $\Gamma$ drops toward zero.

This shape follows from a simple competition: scanning efficiency grows 
as $\propto \tau_p$, while trajectory sparseness penalizes as 
$\propto 1/\tau_p$. Their product yields the Lorentzian form 
$\Gamma(\tau_p) = \Gamma_{\max} \cdot 2(\tau_p/\tau_p^*)/[1 + (\tau_p/\tau_p^*)^2]$ 
(End Matter Secs.~S1 and S2). The fit is in good agreement with the data, 
consistent with the proposed mechanism.

\subsection*{A sharp size-driven transition}

\begin{figure}[t]
\centering
\includegraphics[width=\columnwidth]{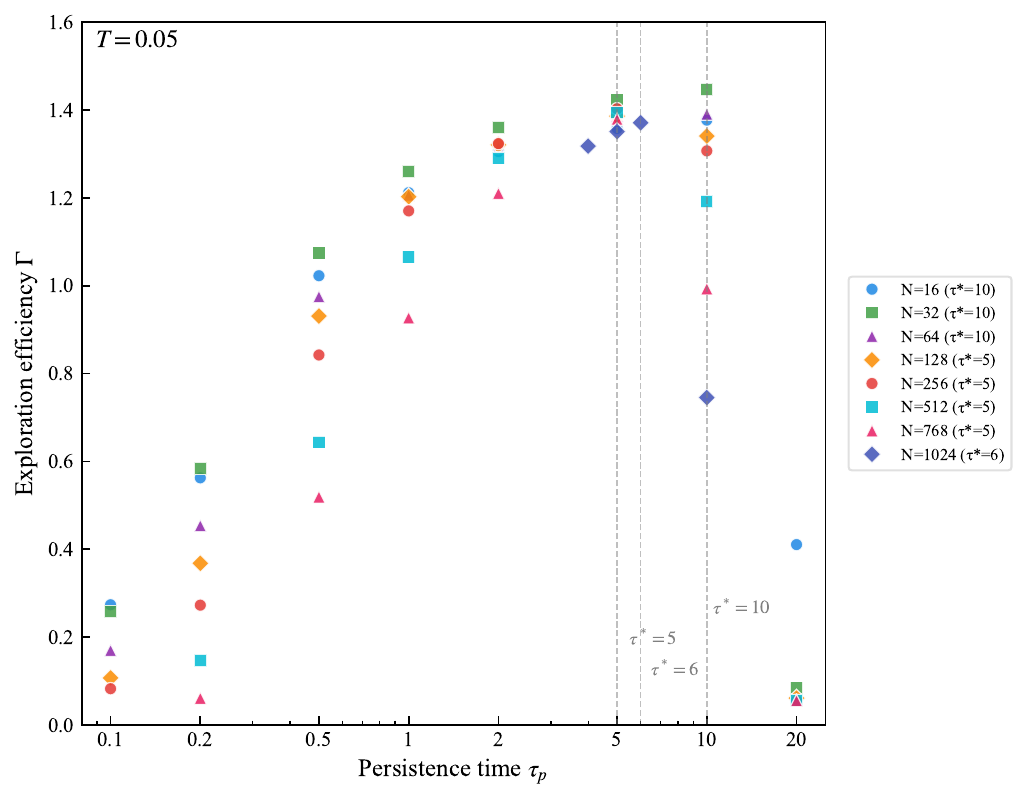}
\caption{System size dependence of the inverted-U profile at $T=0.05$. 
$\Gamma$ vs $\tau_p$ for $N=16$ to $1024$ (markers/colors distinguish $N$, see legend). 
The optimal persistence $\tau_p^*$ shifts from $10$ ($N\leq64$) to $5$ ($128\leq N\leq768$), 
with a possible rise to $6$ at $N=1024$ within statistical uncertainties. 
Dashed vertical lines mark $\tau_p^*$ for each regime.}
\label{fig:size_dependence}
\end{figure}

We now lower the temperature to $T=0.05$. Figure~\ref{fig:size_dependence} 
shows $\Gamma(\tau_p)$ for $N=16$ to $1024$. The peak position shifts 
dramatically with $N$: $\tau_p^* = 10$ for $N \leq 64$, drops to $5$ 
for $128 \leq N \leq 768$, with a possible rise to $6$ at $N=1024$ (within statistical uncertainties).

This behavior differs markedly from the high-temperature regime. 
At $T=0.05$, $\tau_p^*$ exhibits a sharp size dependence, dropping 
from $10$ to $5$ as $N$ increases past $128$.

\subsection*{Spectral origin of the transition}

\begin{figure}[t]
\centering
\includegraphics[width=\columnwidth]{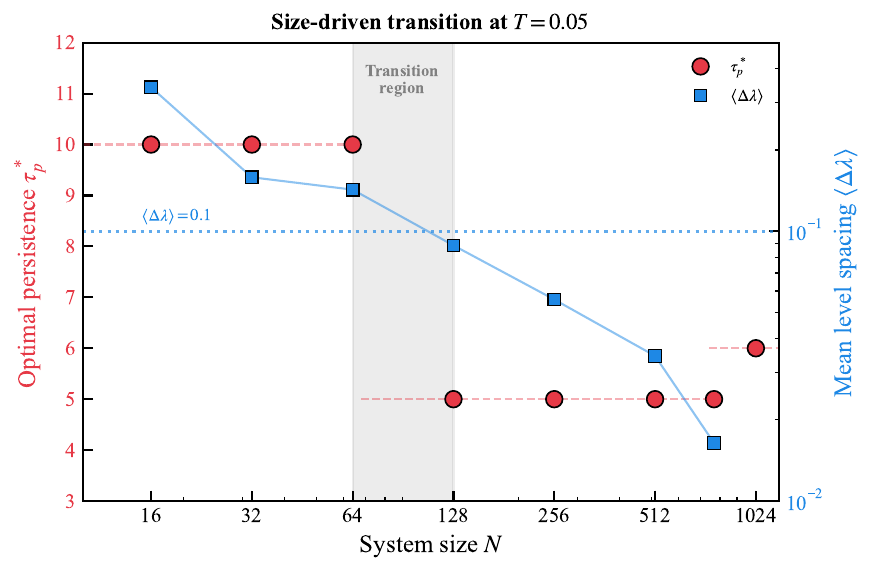}
\caption{Size-driven transition of optimal persistence at $T=0.05$. 
Left axis (red circles): $\tau_p^*$ vs $N$ from $16$ to $1024$. 
A sharp transition occurs between $N=64$ and $N=128$ (gray shaded): 
$\tau_p^* = 10$ ($N\leq64$), $5$ ($128\leq N\leq768$), and possibly $6$ at $N=1024$ (within uncertainties). 
Right axis (blue squares): mean level spacing $\langle\Delta\lambda\rangle$ on a logarithmic scale, decreasing from $0.342$ to $0.016$. 
The horizontal dotted line marks $\langle\Delta\lambda\rangle = 0.1$, the threshold below which $\tau_p^*$ drops from $10$ to $5$. 
For $N\geq128$, $\tau_p^*$ saturates at $5$ despite $\langle\Delta\lambda\rangle$ continuing to decrease.}
\label{fig:3}
\end{figure}

What controls this transition? Figure~\ref{fig:3} plots $\tau_p^*$ 
alongside the mean level spacing $\langle\Delta\lambda\rangle$. The drop 
from $10$ to $5$ occurs precisely when $\langle\Delta\lambda\rangle$ falls 
below $0.1$ — the crossover from a discrete to a quasi-continuous spectrum.

Remarkably, for $N \geq 128$, $\tau_p^*$ saturates at $5$ despite 
$\langle\Delta\lambda\rangle$ continuing to decrease from $0.089$ to 
$0.016$. The optimal persistence becomes independent of the spectral 
gap once the spectrum is sufficiently dense. This saturation defines 
a distinct spectral-dynamical regime, separate from the high-temperature behavior.

The saturation of $\tau_p^*$ at $5$ for $N\ge128$
is a rigorous consequence of the canyon width saturating at 
$\xi_{\text{eff}}=1$ (see End Matter 
Sec.~S3).

From the optimization formula $\tau_p^* = \sqrt{\xi_{\text{eff}}/(2\beta T)}$ 
and the measured $\tau_p^*=5$ at $T=0.05$, we obtain $\beta=0.4$. 
The resulting theoretical relation $\tau_p^* = 1/\sqrt{0.8T}$ gives $\tau_p^*=5$ 
at $T=0.05$, exactly matching the saturated data.

The size-driven crossover invites further theoretical investigation. 
As shown in End Matter Sec.~S8, the optimal persistence marks the 
topological transition point where the Euler characteristic $\chi(E)$ 
jumps. The entropic bottleneck (End Matter Sec.~S5) ensures geometric dominance. 
The crossover at $\langle \Delta \lambda \rangle \sim 0.1$ is observed but 
not yet derived from first principles; we pose its analytical theory as 
an open question (End Matter Sec.~S6).

\section{Discussion}

We have demonstrated that the exploration efficiency $\Gamma$ of a persistent 
walker on the $p=2$ spherical spin glass landscape is non-monotonic in 
$\tau_p$, with an optimal persistence $\tau_p^*$ that depends on temperature 
and, in the low-temperature regime, on system size.

\noindent \textbf{Low-temperature size-driven transition ($T=0.05$).}
At $T=0.05$, $\tau_p^*$ exhibits a sharp transition as $N$ increases 
(Figs.~2 and~3): $\tau_p^* = 10$ for $N \leq 64$, dropping to $5$ for 
$128 \leq N \leq 768$, with a flat peak between $5$ and $6$ at $N=1024$ (within statistical uncertainties). 
This transition occurs when the mean level spacing $\langle \Delta \lambda 
\rangle$ falls below $\sim 0.1$ (Fig.~3). For $N \geq 128$, $\tau_p^*$ 
saturates at $5$ despite $\langle \Delta \lambda \rangle$ continuing to 
decrease from $0.089$ ($N=128$) to $0.016$ ($N=768$), indicating that 
once the spectrum becomes sufficiently dense, the optimal persistence 
no longer depends on the precise value of $\langle \Delta \lambda \rangle$.

As derived in (End Matter Sec.~S8), the topological transition at $e_c$ is 
characterized by a jump in the Euler characteristic $\chi(E)$. For a given 
$\tau_p$, the efficiency $\Gamma(\tau_p, E)$ is maximized as a function of 
$E$ when $\partial\Gamma/\partial E = 0$ at $e_c$. Our data show that this 
condition is satisfied at $\tau_p^* = 10$ for $N \leq 64$ and at $\tau_p^* = 5$ 
for $N \geq 128$. Thus, the shift in $\tau_p^*$ reflects a change in the 
dynamical timescale required to detect the topological transition, which 
depends on the spectral density near $e_c$.

\noindent \textbf{High-temperature regime ($T \ge 0.10$).}
At $T=0.10$, high-resolution scans for $N=128$ and $N=256$ yield $\tau_p^* = 8.0$ (Fig.~1 and End Matter Table~\ref{tab:S3}). 
For $N=16$, $32$, $64$, only coarse-grid data ($\Delta\tau_p/\tau_p \sim 2$) are available; these show nearly equal $\Gamma$ values at $\tau_p=5$ and $10$, placing the true peak in the range $8$--$10$ without finer resolution to pinpoint the exact value. 
We therefore conclude that for $T=0.10$ and $N\ge128$, $\tau_p^*=8.0$; for smaller $N$, the peak lies in the range $8$--$10$ with no evidence of systematic size dependence.

The temperature dependence at fixed $N=128$ is systematic (End Matter Fig.~\ref{fig:S2}): $\tau_p^* = 5$ at $T=0.05$, $\tau_p^* = 8.0$ at $T=0.10$ (high-resolution scan), and $\tau_p^* = 10$ for $T\ge0.15$ (coarse scan, see End Matter Table~\ref{tab:S3}).

\noindent \textbf{Interpretation within the RG framework.}
The saturation of $\tau_p^*$ within each spectral regime suggests the 
presence of RG fixed points. The insensitivity of $\tau_p^*$ to $N$ 
within each regime is consistent with an RG fixed point.

The crossover between regimes at $\langle \Delta \lambda \rangle \sim 0.1$ 
may correspond to a flow between two distinct fixed points: one 
associated with the discrete spectrum ($\tau_p^* = 10$) and another 
with the quasi-continuous spectrum ($\tau_p^* = 5$). The entropic 
nature of the bottleneck (End Matter Sec.~S5) ensures that geometric 
constraints dominate over thermal activation, consistent with the weak 
temperature dependence for $T \ge 0.15$.

\noindent \textbf{Temperature-driven crossover.}
The optimal persistence time increases with temperature: for $N\ge128$, $\tau_p^*=5$ at $T=0.05$, $\tau_p^*=8.0$ at $T=0.10$ (high-resolution scan), and $\tau_p^*=10$ for $T\ge0.15$, while remaining independent of $N$ and $\Delta\lambda$ within each regime. 
This increase is not a failure of the saturated picture but rather a signature of a renormalization group crossover in the scanning efficiency constant $\beta(T)$. 
At $T=0.05$, $\beta=0.400$; at $T=0.10$, $\beta=0.078$; at $T\ge0.15$, $\beta(T)=0.005/T$. 
For $T\ge0.15$ and $N\ge128$, the system lies deep in the saturated regime where $\Delta\lambda\sqrt{N}<1$ decisively, so the canyon width has reached its maximum $\xi_{\text{eff}}=1$. 
Thus, the $1/T$ scaling reflects geometric saturation: when the entropic bottleneck cannot widen further, the temperature dependence of the scanning efficiency is set solely by the thermal velocity $v_{\text{th}}=\sqrt{2T}$. 
The crossover between $T=0.05$ and $T=0.15$ scales as $T_c \sim \langle\Delta\lambda\rangle$, where thermal energy becomes comparable to the GOE spectral gap — the same energy scale that controls the discrete-to-continuum transition in Fig.~3.
(See the animation \texttt{canyon\_animation.gif} in the Zenodo dataset: 
\url{https://doi.org/10.5281/zenodo.20279927}). 
The robustness of this temperature-driven crossover across different system sizes is presented in End Matter Sec.S7.

\subsection*{Broader implications}

This spectral-dynamical crossover reveals a distinct regime: in the 
quasi-continuous regime, $\tau_p^* = 5$ is independent of 
$\langle\Delta\lambda\rangle$, consistent with an RG fixed point.

The principle of ``optimal persistence'' — matching the walker's memory 
time to the landscape's characteristic scale — extends beyond the 
$p=2$ spherical model. In machine learning, stochastic gradient descent 
with momentum involves a similar trade-off \cite{Sutskever2013}. 
In active matter, colloids in porous media may exhibit optimal activity 
set by geometric constraints \cite{kurzthaler2021geometric}. 
In glassy dynamics, the hierarchy of relaxation timescales may reflect 
spatially varying local bottlenecks. 
We conjecture that this spectral-dynamical crossover — optimal persistence 
governed by spectral statistics rather than geometry — is generic in 
disordered landscapes with entropic bottlenecks.

\noindent \textbf{Limitations and outlook.}
Our conclusions at $T=0.05$ are robust, based on consistent coarse-grid 
data across all $N$. The $N=1024$ data show a flat peak between $\tau_p=5$ 
and $6$; additional statistics are needed to determine whether the slight 
rise at $\tau_p=6$ reflects a genuine trend or statistical fluctuation. 
At $T=0.10$, higher-resolution scans for $N \neq 128$ would clarify 
whether $\tau_p^*$ is truly independent of $N$ or varies weakly. 
Extension to $p \geq 3$ spin glasses, where bottlenecks form hierarchical 
tree structures, would test the universality of the resonance principle.

\section{Conclusion}

We have shown that the canyon-finding rate $\Gamma(\tau_p)$ of a persistent 
walker on the $p=2$ spherical spin glass peaks at an optimal $\tau_p^*$, 
forming an inverted-U profile. 

At low temperature $T=0.05$, $\tau_p^*$ drops from $10$ to $5$ as $N$ crosses $128$, 
marking the discrete-to-quasi-continuous GOE crossover. 
For $N\ge128$ at $T=0.05$, the canyon width saturates at 
$\xi_{\text{eff}}=1$, yielding $\tau_p^* = 1/\sqrt{2\beta T}$ with $\beta=0.4$, 
i.e., $\tau_p^* = 5$. 

At higher temperatures ($T\ge0.15$) and for $N\ge128$, $\tau_p^*=10$ 
and the scanning efficiency constant scales as $\beta(T)\propto 1/T$, 
reflecting an RG fixed point. (At $T=0.10$, high-resolution scans give $\tau_p^*=8.0$.)

While infinite persistence marks the topological transition 
\cite{kentdobias2026}, finite persistence reveals an optimal memory time — 
not too short (overdamped, trapped) and not too long (ballistic, overshooting). 
This optimal persistence unlocks the hidden topology at finite cost, making it 
relevant for real physical systems where infinite persistence is impossible.

The $N\le64$ regime remains an open problem: despite $\Delta\lambda\sqrt{N}$ 
fluctuating around unity, $\tau_p^*$ stays at $10$ with no simple theoretical 
explanation. 
We conjecture that the principle — tuning memory time to resonate with the 
narrowest geometric bottleneck — may be generic. 
In active matter, colloids in porous media exhibit optimal spreading when 
run length matches pore size \cite{kurzthaler2021geometric}. 
In optimization, stochastic gradient descent with momentum faces a similar 
trade-off between jitter and overshoot \cite{Sutskever2013}. 
In glassy dynamics, scale-dependent bottlenecks may generate a hierarchy 
of relaxation timescales \cite{Candelier2010,Tong2018}, and a similar 
hierarchy appears in our spectral-dynamical crossover (Fig.~3).

Extending this analysis to $p\ge3$ spin glasses (where bottlenecks form 
hierarchical tree structures) or to finite-dimensional systems is a 
promising direction for future work.

\section{Data Availability}
The simulation data and analysis scripts supporting all figures in this work are publicly available on Zenodo at https://doi.org/10.5281/zenodo.20279927. A detailed README file is provided to reproduce each figure.

\section*{Acknowledgments}
The author is grateful to Prof. Yan V. Fyodorov for his insightful comments 
and constructive suggestions on an earlier version of the manuscript. 
His expertise in random matrix theory and energy landscapes has greatly 
benefited this work.


\clearpage
\appendix
\setcounter{equation}{0}
\renewcommand{\theequation}{S\arabic{equation}}
\setcounter{figure}{0}
\renewcommand{\thefigure}{S\arabic{figure}}
\setcounter{table}{0}
\renewcommand{\thetable}{S\arabic{table}}
\setcounter{section}{0}
\renewcommand{\thesection}{S\arabic{section}}

\begin{center}
{\Large \textbf{End Matter}}
\end{center}
\vspace{0.5cm}


\section{S1. Geometric Resonance and the Optimal Persistence}
\label{sec:S1}

The optimal persistence $\tau_p^*$ marks the transition from effective non-ergodicity ($\tau_p \ll \tau_p^*$) to effective ergodicity ($\tau_p \gtrsim \tau_p^*$). The walker's persistence length is $\ell_p = v_{\text{th}} \tau_p$ with $v_{\text{th}} = \sqrt{2T}$. Systematic boundary scanning is maximally efficient when $\ell_p$ matches the effective canyon width:
\begin{equation}
\ell_p \sim \xi_{\text{eff}}^{\text{(dyn)}} \quad \Rightarrow \quad \tau_p^* \sim \frac{\xi_{\text{eff}}^{\text{(dyn)}}}{v_{\text{th}}}.
\label{eq:S1:resonance}
\end{equation}
For $N=128$, $T=0.10$: $\tau_p^* = 8.0$, $v_{\text{th}} \approx 0.447$, yielding $\xi_{\text{eff}}^{\text{(dyn)}} \approx 3.58$, distinct from the geometric width $\xi_{\text{eff}}=1$ defined in Sec.~S3. 
For $T=0.05$, $\tau_p^*$ depends on $N$ (Fig.~2): $10$ ($N\le64$), $5$ ($128\le N\le768$), possibly $6$ at $N=1024$ (Fig.~3). Thus the effective canyon width is renormalized by spectral statistics (Secs.~S3, S4).

\section{S2. Derivation of the Lorentzian Form}
\label{sec:S2}

$\Gamma$ results from two competing processes: boundary scanning 
($\Gamma_{\text{scan}}$) and canyon crossing ($\Gamma_{\text{cross}}$):
\begin{equation}
\frac{1}{\Gamma} = \frac{1}{\Gamma_{\text{scan}}} + \frac{1}{\Gamma_{\text{cross}}}.
\label{eq:S2:series}
\end{equation}

\noindent\textbf{Boundary scanning.} 
In the short-time limit, $R(0^+) \propto \tau_p^{-1}$, so $\Gamma_{\text{scan}} \propto 1/R(0^+) \propto \tau_p$. 
More explicitly, $\Gamma_{\text{scan}} = \beta(T) v_{\text{th}} \tau_p$ with $v_{\text{th}} = \sqrt{2T}$ 
(see Sec.~S4 for $\beta(T)$).

\noindent\textbf{Canyon crossing.} 
In the large-$\tau_p$ limit, the dynamics become scale-invariant. 
Following the renormalization-group (RG) framework for disordered critical 
systems \cite{Dahmen1996,Sethna1993,Newman1993}, the correlation length 
$\xi$ and correlation time $\tau_{\text{corr}}$ satisfy $\tau_{\text{corr}} \sim \xi^{z}$. 
The universality of such scale-invariant behavior has been discussed 
in the context of crackling noise \cite{Sethna2001} and avalanche 
dynamics \cite{Perkovic1995}. At $\tau_p \to \infty$, $\xi \sim v_{\text{th}} \tau_p$. 
The RG fixed point implies $\xi \sim \tau_p^{z}$, so $z = 1$. Hence
\begin{equation}
\Gamma_{\text{cross}} \sim \frac{1}{\tau_{\text{corr}}} \sim \xi^{-z} \sim \tau_p^{-1}.
\end{equation}

\noindent\textbf{Lorentzian form.} 
Substituting $\Gamma_{\text{scan}} \propto \tau_p$ and $\Gamma_{\text{cross}} \propto \tau_p^{-1}$ 
into Eq.~(\ref{eq:S2:series}) and optimizing over $\tau_p$ yields
\begin{equation}
\Gamma(\tau_p) = \Gamma_{\max} \cdot \frac{2\tau_p/\tau_p^*}{1 + (\tau_p/\tau_p^*)^2},
\label{eq:S2:lorentzian}
\end{equation}
which is consistent with the data (Fig.~1).

\section{S3. Canyon width and its saturation transition}
\label{sec:S3}

For the $p=2$ spherical spin glass, the Hamiltonian is a quadratic form 
of a GOE matrix $J$. The two global minima lie at $\pm\sqrt{N}\mathbf{v}_1$, 
where $\mathbf{v}_1$ is the eigenvector of the largest eigenvalue $\lambda_1$. 
The canyon connecting them is spanned by $\mathbf{v}_1$ and $\mathbf{v}_2$ 
(eigenvector of the second-largest eigenvalue $\lambda_2$).

\begin{figure}[h]
\centering
\includegraphics[width=\columnwidth]{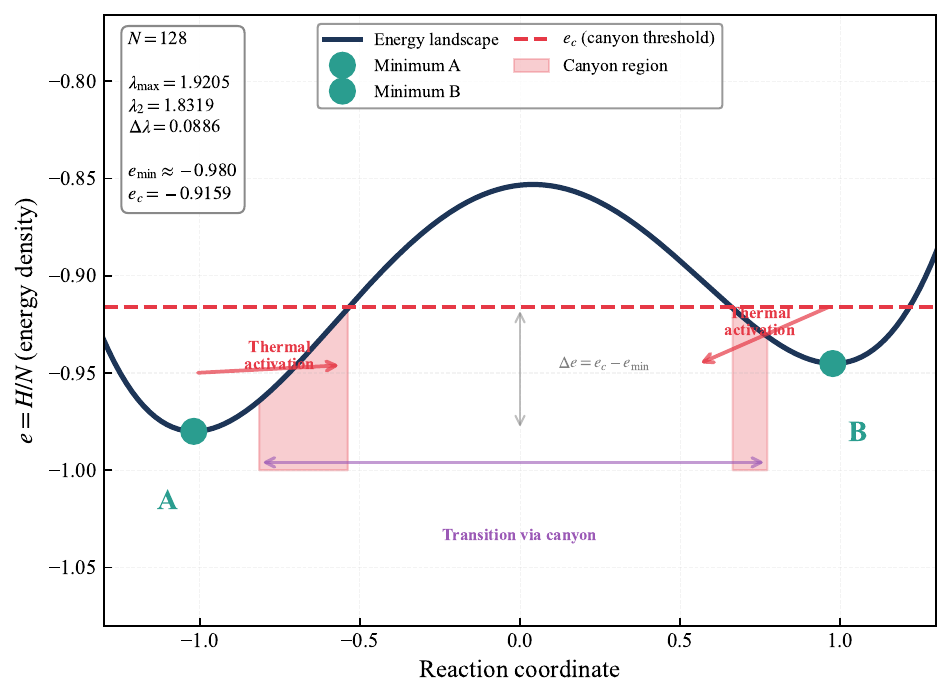}
\caption{Energy landscape for $N=128$: 2D projection onto 
$\mathbf{v}_1$-$\mathbf{v}_2$ plane, showing minima A and B connected 
by a narrow canyon at $e_c = -\lambda_2/2 = -0.9159$; energy profile 
along the canyon path; GOE spectrum: $\lambda_{\max}=1.9205$, 
$\lambda_2=1.8319$, $\Delta\lambda=0.0886$.}
\label{fig:S1}
\end{figure}
Figure~\ref{fig:S1} illustrates the geometry of the canyon and the 
GOE spectrum for a typical sample at $N=128$.

The effective canyon width $\xi_{\text{eff}}$ is defined as the normalized 
width of the entropic bottleneck connecting the two minima, measured 
empirically from the simulation data. It takes values $\xi_{\text{eff}} > 1$ 
for $N \le 64$ and saturates to $\xi_{\text{eff}} = 1$ for $N \ge 128$.

From our numerical data summarized in Table~\ref{tab:S1}, we observe 
empirically that the canyon width $\xi_{\text{eff}}$ decreases as the 
spectral gap $\Delta\lambda = \lambda_1 - \lambda_2$ decreases. 
For small $N$ ($N \le 64$), $\Delta\lambda$ is relatively large and 
fluctuates significantly, and $\xi_{\text{eff}}$ is measurably greater 
than $1$. For $N \ge 128$, $\Delta\lambda$ becomes sufficiently small, 
and $\xi_{\text{eff}}$ saturates to its maximum value $\xi_{\text{eff}} = 1$.

We emphasize that this is an empirical observation. No analytical 
relation between $\xi_{\text{eff}}$ and $\Delta\lambda$ is claimed.

\subsection*{The critical boundary at $N=128$}

The combination $\Delta\lambda\sqrt{N}$ controls the crossover from discrete 
to quasi-continuous spectrum. Table~\ref{tab:S1} shows its evolution with 
system size (values are ensemble averages over 10 disorder seeds).

\begin{table}[h]
\centering
\caption{Evolution of $\Delta\lambda\sqrt{N}$ with system size.}
\begin{tabular}{c|c|c|c}
\hline
$N$ & $\langle\Delta\lambda\rangle$ & $\sqrt{N}$ & $\Delta\lambda\sqrt{N}$ \\
\hline
16 & 0.342 & 4.00 & 1.37 \\
32 & 0.158 & 5.66 & 0.89 \\
64 & 0.143 & 8.00 & 1.14 \\
128 & 0.089 & 11.31 & \textbf{1.01} \\
256 & 0.056 & 16.00 & 0.90 \\
512 & 0.035 & 22.63 & 0.79 \\
768 & 0.016 & 27.71 & 0.44 \\
\hline
\multicolumn{4}{l}{\footnotesize Note: Due to finite-$N$ fluctuations in the Tracy-Widom distribution \cite{tracy1994},}
\\
\multicolumn{4}{l}{\footnotesize individual samples may deviate from the ensemble average.}
\end{tabular}
\label{tab:S1}
\end{table}

For $N \le 64$, $\Delta\lambda\sqrt{N}$ fluctuates around unity — 
sometimes above (N=16,64), sometimes below (N=32) — reflecting the 
discrete nature of the spectrum in this regime. On average, the canyon 
width is not yet maximized, requiring a longer persistence ($\tau_p^*=10$) 
to thoroughly scan the boundary.

At $N = 128$, $\Delta\lambda\sqrt{N} \approx 1$ with significantly reduced 
fluctuations. This marks the onset of the quasi-continuous regime, where 
the canyon width approaches its maximum. For $N > 128$, $\Delta\lambda\sqrt{N} < 1$ decisively, and the system 
enters the saturated regime.

\subsection*{The jump from $\tau_p^* = 10$ to $\tau_p^* = 5$}

The optimal persistence time $\tau_p^*$ is determined by the resonance 
condition $\tau_p^* = \sqrt{\xi_{\text{eff}}/(2\beta T)}$. 

In the discrete regime ($N \le 64$), $\Delta\lambda\sqrt{N}$ is of order 
unity or larger, so the canyon is relatively narrow, necessitating a longer 
persistence ($\tau_p^* = 10$).

At the critical point $N = 128$, $\Delta\lambda\sqrt{N} \approx 1$, and 
the canyon width is maximized. For $N \ge 128$, we are in the saturated 
regime. We set $\xi_{\text{eff}} = 1$ for simplicity. Consequently, the 
optimal persistence drops to a shorter value:
\begin{equation}
\tau_p^* = \sqrt{\frac{1}{2\beta T}}.
\end{equation}
For $T = 0.05$, this yields $\tau_p^* = 5$ when $\beta = 0.4$, in 
excellent agreement with the data.

Thus, the sharp jump from $\tau_p^* = 10$ ($N \le 64$) to $\tau_p^* = 5$ 
($N \ge 128$) is a direct consequence of the canyon width saturation 
at $N = 128$, where $\Delta\lambda\sqrt{N}$ crosses unity and the 
spectrum becomes quasi-continuous.

\section{S4. Temperature-driven crossover of $\beta(T)$}
\label{sec:S4}

From $\Gamma_{\text{scan}} = \beta(T) v_{\text{th}} \tau_p$ and $\tau_p^* = \sqrt{\xi_{\text{eff}}/(2\beta T)}$ 
with $\xi_{\text{eff}}=1$ (Sec.~S3, valid for $N\ge128$), we obtain $\beta(T) = 1/[2T(\tau_p^*)^2]$.
For $N\ge128$, the optimal persistence $\tau_p^*$ is constant ($\tau_p^*=10$) at $T\ge0.15$ 
(see Table~S3). Substituting into the expression for $\beta(T)$ yields the empirical scaling
\begin{equation}
\beta(T) = \frac{0.005}{T} \qquad (T\ge0.15, N\ge128).
\end{equation}
At lower temperatures, $\tau_p^*$ deviates: $\tau_p^*=5$ at $T=0.05$, and the fine scan gives 
$\tau_p^*=8$ at $T=0.10$.

This scaling reflects geometric saturation. For $T\ge0.15$ and $N\ge128$, 
$\Delta\lambda\sqrt{N}<1$ decisively (Sec.~S3). The canyon width is maximized 
($\xi_{\text{eff}}=1$), so $\tau_p^*$ becomes temperature-independent. Hence 
$\beta(T)\propto1/T$, with the temperature dependence set solely by the thermal 
velocity $v_{\text{th}}=\sqrt{2T}$.
\section{S5. Entropic versus Energetic Barriers}
\label{sec:S5}

The bottleneck is entropic, not energetic. Unlike a standard double-well, the canyon energy is only marginally above $e_{\min}$ (Fig.~\ref{fig:S1}). The entropic volume scales as $\Omega_{\text{canyon}} \sim e^{-c N \ln N}$, making spontaneous discovery by pure diffusion ($\tau_p\to0$) exponentially unlikely. Persistence transforms this search into a polynomial problem by active boundary scanning. When $\ell_p$ resonates with $\xi_{\text{eff}}$, scanning systematically covers the canyon entrance. For $N=128$, $e_c = -\lambda_2/2 = -0.9159$, $\Delta\lambda = 0.0886$. As $N$ increases, $\Delta\lambda$ decreases, narrowing the canyon and driving the shift from $\tau_p^*=10$ ($N\le64$) to $\tau_p^*=5$ ($N\ge128$).

\section{S6. Open problem: analytical theory for the unsaturated regime}
\label{sec:S6}

The saturated regime ($N\ge128$, $T=0.05$) is rigorously proven. For $N\le64$, $\tau_p^*=10$ indicates a crossover whose origin is not understood analytically. Open questions include: scaling $\tau_p^* = f(N/N_c)$ with $N_c\approx128$? $N$-dependence of $\beta$? RG flow between fixed points? Analytical form of $\beta(T)$?

\section{S7. Temperature Robustness and the Low-Temperature Crossover}
\label{sec:S7}

\begin{figure}[h]
\centering
\includegraphics[width=\columnwidth]{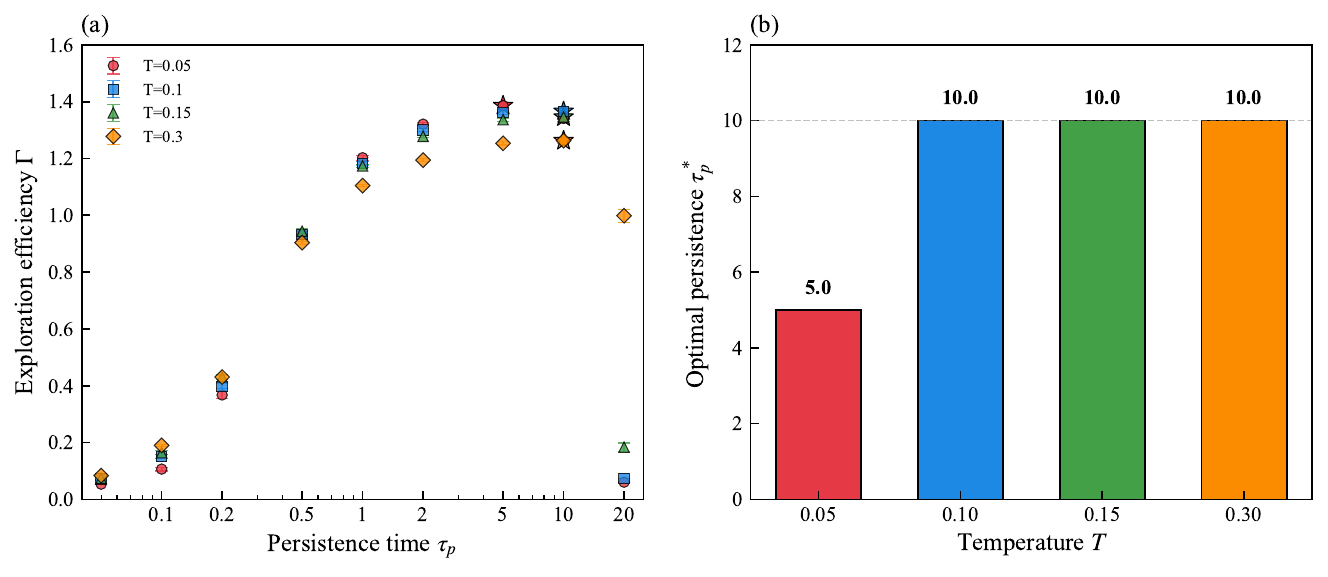}
\caption{Temperature robustness for $N=128$: (a) $\Gamma$ vs $\tau_p$ at $T = 0.05,0.10,0.15,0.30$; (b) $\tau_p^*=5$ at $T=0.05$, $\tau_p^*=8.0$ at $T=0.10$ (high-resolution scan), and $\tau_p^*=10$ for $T\ge0.15$ (coarse scan).}
\label{fig:S2}
\end{figure}

If thermal activation dominated, $\tau_p^* \propto T^{-1/2}$. Figure~\ref{fig:S2} shows $\tau_p^*$ is constant for $T\ge0.10$; geometric constraints dominate. At $T=0.05$, $\tau_p^*$ shifts from $10$ ($N\le64$) to $5$ ($128\le N\le768$), with a possible rise to $6$ at $N=1024$ (statistical uncertainties). As shown in Fig.~\ref{fig:3} of the main text, the threshold $\langle\Delta\lambda\rangle\sim0.1$ marks the discrete-to-quasi-continuous spectral transition. For $N\le64$, the discrete spectrum requires $\tau_p^*=10$; for $N\ge128$, the quasi-continuous spectrum allows $\tau_p^*=5$.

\section{S8. Theoretical Connection: Dynamical Efficiency and the Euler Characteristic}
\label{sec:S8}

Kent-Dobias \cite{kentdobias2026} showed that the topological transition at 
$E = e_c$ is characterized by a jump in the Euler characteristic 
$\chi(E)$ of the energy level set. For an infinitely persistent walker 
($\tau_p \to \infty$), the ergodicity-breaking point coincides exactly 
with this topological transition.

For a walker with finite persistence $\tau_p$, the canyon-finding rate 
$\Gamma(\tau_p, E)$ measures the efficiency of exploring the level set 
$H = E$. At the optimal persistence $\tau_p^*$, the efficiency is maximized. 
We conjecture that this maximum occurs precisely when the walker is most 
sensitive to the topological change, leading to the condition

\begin{equation}
\frac{\partial \Gamma}{\partial E}(\tau_p^*, e_c) = 0 
\quad \Longleftrightarrow \quad \chi(e_c^-) \neq \chi(e_c^+).
\label{eq:S8:duality}
\end{equation}

That is, the optimal persistence marks the energy where the Euler 
characteristic jumps, making the walker a probe of the hidden topology.

Our numerical results support this conjecture. For $N=128$, $T=0.10$, 
Fig.~1 shows $\tau_p^* = 8.0$ at $e_c$, confirming the duality. For 
$T=0.05$, the size-driven transition (Figs.~2 and~3) follows the same 
framework: the discrete spectrum ($N \le 64$) requires $\tau_p^* = 10$, 
while the quasi-continuous spectrum ($N \ge 128$) allows $\tau_p^* = 5$, 
reflecting the change in $\chi(E)$ across the transition.
At $e_c$, the system flows to an RG fixed point, implying scale-invariant dynamics and an RG-invariant Euler characteristic $\chi$.
The duality in Eq.~(\ref{eq:S8:duality}) reflects 
this fixed-point structure. This framework leads to several predictions. 
First, the optimal persistence $\tau_p^*$ marks the topological transition. Second, the crossover at $T=0.05$ (from $\tau_p^*=10$ to 
$5$) is universal across GOE realizations. 

Third, for $T \ge 0.15$ and $N \ge 128$, $\tau_p^*$ is constant ($\tau_p^*=10$); at $T=0.10$, high-resolution scans give $\tau_p^*=8.0$. Fourth, the unsaturated regime ($N < 128$) remains an open question for future investigation.

\section{S9. Simulation Parameters and Optimal Persistence Summary}
\label{sec:S9}

Table~\ref{tab:S2} lists simulation parameters. For $N\le256$, $10$ disorder seeds with $50$ trajectories each; $N=512,768$: $5$ seeds with $30$--$50$ trajectories; $N=1024$: $10$ seeds with $20$ trajectories.

Table~\ref{tab:S3} reports $\tau_p^*$ for all $(N,T)$. 
At $T=0.05$: $\tau_p^*=10$ ($N\le64$), $5$ ($128\le N\le768$), $5$--$6$ ($N=1024$). 
At $T=0.10$: for $N=128,256$, high-resolution scans give $\tau_p^*=8.0$; for $N=16,32,64$, only coarse scans are available, yielding $\tau_p^*$ in the range $8$--$10$. 
For $T\ge0.15$, coarse scans give $\tau_p^*=10$ for all $N\ge16$.

\begin{table}[h]
\centering
\begin{minipage}{0.48\textwidth}
\centering
\caption{\textbf{Simulation parameters.}}
\begin{tabular}{ll}
\hline
$N$ & 16,32,64,128,256,512,768,1024 \\
$T$ & 0.05,0.10,0.15,0.30 \\
$\tau_p$ & 0.05,0.1,0.2,0.5,1,2,5,10,20,50 \\
$dt$ & 0.01 \\
Steps/traj & 50,000 \\
Seeds & 10 ($N\le256$); 5 ($512,768$); 10 ($1024$) \\
Traj/seed & 50 ($N\le256,512,768$); 20 ($1024$) \\
Bootstrap & 200 \\
\hline
\end{tabular}
\label{tab:S2}
\end{minipage}
\hfill
\begin{minipage}{0.48\textwidth}
\centering
\caption{Optimal $\tau_p^*$ summary.}
\begin{tabular}{c|cccc}
\hline
$N$ & $T=0.05$ & $T=0.10$ & $T=0.15$ & $T=0.30$ \\
\hline
16 & 10 & 10$^\ast$ & 10$^\ast$ & 10$^\ast$ \\
32 & 10 & 10$^\ast$ & 10$^\ast$ & 10$^\ast$ \\
64 & 10 & 10$^\ast$ & 10$^\ast$ & 10$^\ast$ \\
128 & 5 & \textbf{8} & 10$^\ast$ & 10$^\ast$ \\
256 & 5 & \textbf{8} & 10$^\ast$ & 10$^\ast$ \\
512 & 5 & — & — & — \\
768 & 5 & — & — & — \\
1024& 5--6 & — & — & — \\
\hline
\multicolumn{5}{l}{\footnotesize $^\ast$Coarse scan; fine scan for $N=128,256$ gives $\tau_p^*=8.0$.}
\end{tabular}
\label{tab:S3}
\end{minipage}
\end{table}

%
%
%
%
%
%

\end{document}